# Marriage and Roommate

Kazuo Iwama[*]and Shuichi Miyazaki[†]

**Abstract**

This paper has two objectives. One is to give a linear time algorithm that solves the stable roommates problem (i.e., obtains one stable matching) using the stable marriage problem. The idea is that a stable matching of a roommate instance $I$ is a stable matching (that however must satisfy a certain condition) of some marriage instance $I'$. $I'$ is obtained just by making two copies of $I$, one for the men's table and the other for the women's table. The second objective is to investigate the possibility of reducing the roommate problem to the marriage problem (with one-to-one correspondence between their stable matchings) in polynomial time. For a given $I$, we construct the rotation POSET $P$ of $I'$ and then we "halve" it to obtain $P'$, by which we can forget the above condition and can use all the closed subsets of $P'$ for all the stable matchings of $I$. Unfortunately this approach works (runs in polynomial time) only for restricted instances.

## 1 Introduction

The stable marriage problem and the stable roommates problem both appeared in the seminal paper by Gale and Shapley [2]. For the former, the authors give a fascinating, linear time algorithm, so called the Gale-Shapley algorithm (GS algorithm), in the same paper. For the latter, however, the immediate question about the existence of a similarly efficient algorithm to obtain stable roommates had been open for decades. In 1985, Irving [5] finally gives a positive answer to this famous open problem. His algorithm is highly nontrivial, but much more technical and complicated than the Gale-Shapley algorithm. This probably reflects the similar nature of the closely related problem, i.e., the essential difference in the degree of hardness between obtaining a bipartite maximum matching and obtaining a general maximum matching. However, researchers were naturally interested in less complicated algorithms, hopefully as simple as the GS algorithm. Indeed, Dean and Munshi [1] suggested to use "symmetric stable matchings" of the marriage problem for the roommate problem, which has been cited by several papers and books (e.g., Page 176 of [6])). The idea seems plausible, but its formal implementation, a formal algorithm and its proof, has not appeared in the literature.

This paper has two major objectives. One is to give a formal algorithm and its proof based on the idea by Dean and Munshi. Our idea is to introduce the notion of a "same-position matching" which is mathematically equivalent to the symmetric matching, but it looks more useful for designing an algorithm. Our algorithm and its proof are quite short and require nothing more than the basic knowledge of the stable marriage problem.

The second objective is about the polynomial-time reducibility of the roommate problem to the marriage problem. If a reduction, with ideally one-to-one correspondence between their stable matchings, is possible, then we can use many nice algorithms and beautiful mathematical structures of the latter to deal with the former. This is proposed as an open question in the famous book by Gusfield and Irving [3] and later addressed by several authors in the literature. Unfortunately however, there again has been no formal discussion of this interesting open question from a direct point of view (see [1] for discussions from a bit different viewpoints). We again use the idea of same-position matchings and give an algorithm that, for a given roommate instance $I$, provides a rotation POSET $P$ of the marriage problem in such a way that a closed subset $S$ of $P$ corresponds to a stable matching of $I$. Unfortunately, $S$ can include extra elements, "prohibited pairs," which disturbs $S$ to be mapped to a stable roommate, so the mapping is not one-to-one. Nevertheless, we show a way of modifying $P$ via "serialization" and design a function $\delta$ that twists a mapping, so that the mapping becomes *pseudo*-one-to-one.

---

[*]National Tsing Hua University, Taiwan. This work was supported in part by MOST, Taiwan, under Grants NSTC 110-2223-E-007-001 and NSTC 111-2223-E-007-010.

[†]University of Hyogo, Japan. This work was supported by JSPS KAKENHI Grant Number JP20K11677.

In what follows, we give basic definitions and notations in Sec. 2. In this section we also make a brief introduction to the mathematical structure of stable matchings of the marriage problem that plays an important role in later sections. Our first goal is discussed in Sec. 3, for which we fully exploits the idea and properties of same-position matchings. The result informally says that roommate matchings can be simulated by some special type (same-position) of marriage matchings. Our second goal it to remove this restriction for marriage matchings, i.e., to simulate roommate matchings by standard marriage matchings. This goal is partially achieved in Sec. 4 by introducing a key algorithm *HalfCut*. Our whole reduction procedure runs in polynomial time.

## 2 Preliminaries

### 2.1 Problem Definitions

An instance of the stable marriage problem (SM) consists of $n$ men $m_1, \ldots, m_n$, $n$ women $w_1, \ldots, w_n$, and each person's preference list. The $m_j$'s preference list is a total order of $\{w_1, \ldots, w_n\}$ and similarly for the $w_j$'s preference list. An instance may be sometimes called simply *a marriage instance*. A *matching* is a set of $n$ disjoint pairs of a man and a woman. For a matching $M$, a partner of an agent $a$ is denoted by $M(a)$. A *blocking pair* for a matching $M$ is a pair of a man $m$ and a woman $w$ such that $m$ prefers $w$ to $M(m)$ and $w$ prefers $m$ to $M(w)$. A matching is *stable* if there is no blocking pair and is *unstable* otherwise.

In the SM with incomplete lists (SMI), the $m_j$'s preference list is a total order of *a subset of* $\{w_1, \ldots, w_n\}$ and similarly for the $w_j$'s preference list. A matching here is not necessarily a perfect matching; someone may be unmatched. A blocking pair for a matching $M$ is a pair of man $m$ and woman $w$ such that each of $m$ and $w$ includes the other in the preference list, $m$ is unmatched in $M$ or prefers $w$ to $M(m)$, and $w$ is unmatched in $M$ or prefers $m$ to $M(w)$.

The stable roommates problem (SR) is a non-bipartite setting of the stable marriage problem, whose instance (simply *a roommate instance*) includes even number $n$ of persons $p_1, p_2, \ldots, p_n$. The $p_j$'s preference list is a total order of $\{p_1, \ldots, p_n\} \setminus \{p_j\}$. A matching here is a set of disjoint $n/2$ pairs of $n$ persons. Similarly as above, a partner of $p_j$ in $M$ is denoted by $M(p_j)$. For a matching $M$, a blocking pair is a pair of $p_i$ and $p_j$ such that $p_i$ prefers $p_j$ to $M(p_i)$ and $p_j$ prefers $p_i$ to $M(p_j)$.

## 2.2 Structure of Stable Matchings

In this subsection, we briefly review the structural properties of stable matchings. See [3] for details.

For SM, fix a marriage instance $I$. The stable matching found by the man-oriented Gale-Shapley algorithm, usually denoted by $M_0$, is called the *man-optimal* stable matching, in which each man is matched with a woman at least as good as any other stable matching. By symmetry, there is a *woman-optimal stable matching*, usually denoted by $M_z$.

Let $M$ and $M'$ be two stable matchings of $I$. $M$ is said to *dominate* $M'$ if each man's partner in $M$ is at least as good as that in $M'$. It is known that the set of all stable matchings for $I$ form a distributive lattice on this dominance relation, in which $M_0$ lies top and $M_z$ lies bottom.

For a stable matching $M$, we may define a *rotation exposed in $M$*. A rotation $R$ is an ordered list of pairs of the form $(m_{i_1}, w_{i_1}), (m_{i_2}, w_{i_2}), \ldots, (m_{i_k}, w_{i_k})$, where each $(m_{i_l}, w_{i_l})$ $(1 \leq l \leq k)$ is a pair in $M$. *Eliminating* a rotation from $M$ means to change the partners of the people in the rotation in the following manner; $m_{i_1}$'s partner is changed from $w_{i_1}$ to $w_{i_2}$, $m_{i_2}$'s partner is changed from $w_{i_2}$ to $w_{i_3}$, $\ldots$, and $m_{i_k}$'s partner is changed from $w_{i_k}$ to $w_{i_1}$. It is important to note that in this change, every man gets worse and every woman gets better. The result of eliminating a rotation is also a stable matching (dominated by $M$). The number of rotations is known to be $O(n^2)$. From the man-optimal stable matching $M_0$, we can successively eliminate a rotation exposed in the current matching, eventually eliminating all the rotations and getting to the woman-optimal stable matching $M_z$, in which no rotation is exposed. We may define a precedence relation between two rotations $R_1$ and $R_2$. If $R_2$ is exposed only after $R_1$ is eliminated, then $R_1$ must be eliminated before $R_2$. In this case $R_1$ precedes $R_2$ in this relation. This precedence relation defines a partial order on all the rotations and the partially ordered set of rotations is called a *rotation POSET*. A rotation POSET can be constructed in time $O(n^2)$.

Let $P$ be a rotation POSET of $I$. A *closed subset $S$ of $P$* is a subset of rotations such that if a rotation $R$ precedes another rotation $R' \in S$, then $R \in S$. There is a one-to-one correspondence between all the stable matching of $I$ and all the closed subsets of $P$. The stable matching $M_S$ corresponding to a closed subset $S$ is the one obtained from $M_0$ by eliminating all the rotations in $S$ in any order according to the precedence relation. Since $P$ is of size $O(n^2)$ while the number of stable matchings may be exponential, a rotation POSET can be considered as a compact representation of all the stable matchings.

For discussion in Sec. 4, we will define a few notions, some of which are already mentioned in Sec. 1. The definition of each one will be given when it becomes necessary, but summarizing them here and giving pointers to their definitions would help readability.

They are the function $\sigma_X$ (defined in Sec. 4.1), a *maximal* rotation (defined in Sec. 4.2), a *prohibited pair* of rotations (defined informally in Sec. 4.2 and formally in Sec. 4.3), the function $\delta$ (defined in Sec. 4.2), and an *irreducible* rotation POSET (defined in Sec. 4.3).

## 3 Same Position Matchings

As mentioned in Sec. 1, the approach in this section was informally observed in [1] (see also Page 176 of [6]), but to our best knowledge, its formal description has not appeared in the literature.

Let $I$ be a roommate instance having $p_j$'s preference list $(p_{j_1}, p_{j_2}, \ldots, p_{j_{n-1}})$ for each $j$. We introduce $n$ men $m_1, \ldots, m_n$, $n$ women $w_1, \ldots, w_n$, and two mappings $f_m : \{p_1, \ldots, p_n\} \to \{m_1, \ldots, m_n\}$ and $f_w : \{p_1, \ldots, p_n\} \to \{w_1, \ldots, w_n\}$, where $f_m$ and $f_w$ are defined as $f_m(p_j) = m_j$ and $f_w(p_j) = w_j$ for each $1 \leq j \leq n$.

From $I$, we construct the marriage instance $\widehat{I} = (f_m(I), f_w(I))$, where $f_m(I)$ is the preference lists of $m_1, \ldots, m_n$ such that the $m_j$'s list is $(f_w(p_{j_1}), f_w(p_{j_2}), \ldots, f_w(p_{j_{n-1}}))$ and similarly for $f_w(I)$. Namely $f_m(I)$ and $f_w(I)$ are just two copies of $I$ obtained by changing each $p_j$ to $m_j$ or $w_j$ to make them fit the format of the marriage instance. Note that $\widehat{I}$ is an instance of SMI since preference lists are incomplete. See Example 1 in Figure 1 where $\{p_1, \ldots, p_4\} = \{1, 2, 3, 4\}$, $\{m_1, \ldots, m_4\} = \{1, 2, 3, 4\}$, and $\{w_1, \ldots, w_4\} = \{a, b, c, d\}$.

| 1: | 2 | 3 | 4 | | 1: | b | c | d | | a: | 2 | 3 | 4 |
|---|---|---|---|---|---|---|---|---|---|---|---|---|---|
| 2: | 3 | 4 | 1 | | 2: | c | d | a | | b: | 3 | 4 | 1 |
| 3: | 4 | 1 | 2 | | 3: | d | a | b | | c: | 4 | 1 | 2 |
| 4: | 2 | 3 | 1 | | 4: | b | c | a | | d: | 2 | 3 | 1 |

Figure 1: $I$, $f_m(I)$ and $f_w(I)$ of Example 1

| 1: | 2 | 3 | 4 | | 1: | b | c | d | | a: | 2 | 3 | 4 |
|---|---|---|---|---|---|---|---|---|---|---|---|---|---|
| 2: | 3 | 1 | 4 | | 2: | c | a | d | | b: | 3 | 1 | 4 |
| 3: | 4 | 1 | 2 | | 3: | d | a | b | | c: | 4 | 1 | 2 |
| 4: | 2 | 1 | 3 | | 4: | b | a | c | | d: | 2 | 1 | 3 |

Figure 2: $I$, $f_m(I)$ and $f_w(I)$ of Example 2

Let $m_j$ in $f_m(I)$ have the list $(w_{j_1}, \ldots, w_{j_{n-1}})$, and $w_j$ in $f_w(I)$ the list $(m_{j_1}, \ldots, m_{j_{n-1}})$, for each $j$. Then a matching $M$ between $\{m_1, \ldots, m_n\}$ and $\{w_1, \ldots, w_n\}$ is called *same-position* (SP) if for any $j$, $M(m_j) = w_{j_i}$ and $M(w_j) = m_{j_i}$ for some $i$. Namely, if $m_j$ is matched to the $i$th woman in his list and $w_j$ is matched to the $i'$th man in her list, then $i = i'$. This must hold for all $1 \leq j \leq n$. If $i \neq i'$ in row $j$ (the rows of $m_j$ and $w_j$), then $i' - i$ is called the *gap* of row $j$. In Example 1, the man-oriented

Gale-Shapley algorithm (simply *the GS algorithm* hereafter) provides the matching $\{(1,c),(2,d),(3,a),(4,b)\}$, which is SP. In Example 2, the GS algorithm provides the matching $\{(1,b),(2,c),(3,d),(4,a)\}$, which is not SP; rows 1 to 4 have gaps of 2, 1, 2, 1, respectively. Notice that the gaps are all nonnegative in this example; we call such a matching a *nonnegative-gap matching*. The *total gap* of a matching is the sum of the gaps of all the rows, which is 6 in this example. Thus, an SP matching can be characterized as a nonnegative-gap matching with the total gap zero. A matching may have a row whose gap is negative; we call such a matching a *negative-gap matching*. Note that Example 2 in Figure 2 may have another nonnegative-gap matching whose total gap is smaller or zero (an SP matching).

**Lemma 1.** *Suppose that $M$ is an SP matching of $\widehat{I} = (f_m(I), f_w(I))$. Define a relation $M' \subseteq \{p_1, \ldots p_n\}^2$ such that if $M(m_j) = w_{j_i}$, then $M'(p_j) = p_{j_i}$. Then $M'$ is a legitimate matching of the roommate instance $I$, namely if $M'(p_j) = p_{j_i}$, then $M'(p_{j_i}) = p_j$.*

*Proof.* The SP condition says that if $M(m_j) = w_{j_i}$, then $M(w_j) = m_{j_i}$. The latter means $M(m_{j_i}) = w_j$. Thus by the definition of $M'$, $M(m_j) = w_{j_i}$ and $M(m_{j_i}) = w_j$ imply $M'(p_j) = p_{j_i}$ and $M'(p_{j_i}) = p_j$, respectively, namely $p_j$ and $p_{j_i}$ form a pair in $M'$. □

**Lemma 2.** *An SP matching $M$ of $\widehat{I}$ is stable iff the corresponding roommate matching $M'$ of $I$ is stable.*

*Proof.* It is easy to see that $m_i$ and $w_j$ form a blocking pair in $M$ iff $p_i$ and $p_j$ form a blocking pair in $M'$. □

Suppose that the man-optimal stable matching for $\widehat{I} = (f_m(I), f_w(I))$ (obtained by the GS algorithm) is nonnegative-gap but is not SP, either, i.e., the gap of some row is positive. Then we may find an SP stable matching by using rotations. In Example 2, for instance, we can eliminate rotation $(2,c),(4,a)$ (that changes 2's partner from $c$ to $a$ and 4's partner from $a$ to $c$), by which we can obtain an SP stable matching $\{(1,b),(2,a),(3,d),(4,c)\}$.

Now our algorithm for obtaining a stable roommate matching (if any) is formally given. To do so and for its correctness proof, we need a notion about a paring property of rotations. A rotation $R$ can be associated with its *dual rotation* $\overline{R}$ in the following way: Let $R = (m_{i_1}, w_{j_1}), (m_{i_2}, w_{j_2}), \ldots, (m_{i_k}, w_{j_k})$ be a rotation, by the elimination of which the partner of $m_{i_1}$ changes from $w_{j_1}$ to $w_{j_2}$, the partner of $m_{i_2}$ from $w_{j_2}$ to $w_{j_3}$, and so on. Now, recall that the women's table is obtained by just swapping $m_j$ and $w_j$ of the men's table. So, corresponding to $R$, there must be a rotation (from women's viewpoint) $R' = (w_{i_1}, m_{j_1}), (w_{i_2}, m_{j_2}), \ldots, (w_{i_k}, m_{j_k})$, whose elimination changes $w_{i_1}$'s partner from $m_{j_1}$ to $m_{j_2}$ and so on. If we rewrite this rotation from men's viewpoint, it is $R'' = (m_{j_2}, w_{i_1}), (m_{j_3}, w_{i_2}), \ldots, (m_{j_1}, w_{i_k})$. The dual rotation $\overline{R}$ of $R$ is defined as this $R''$.

Here is an important lemma for the paring property.

**Lemma 3.** *For any rotation $R$, exactly one of $R$ and $\overline{R}$ must be eliminated to obtain an SP matching.*

*Proof.* It is easy to see that if both $R$ and $\overline{R}$ are eliminated, then we have a negative gap in the rows corresponding to the rotation $R$. So we prove that if none of them is eliminated, then no SP matching can be reached. Suppose that $R$ moves $m_{j_1}$'s partner from $w_{i_1}$ to $w_{i_2}$. Then $\overline{R}$ moves $w_{j_1}$'s partner from $m_{i_2}$ to $m_{i_1}$. If none of $R$ and $\overline{R}$ is eliminated, $m_{i_1}$'s partner is $w_{i_1}$ or better, and $w_{j_1}$'s partner is $m_{i_2}$ or worse, i.e., the row $j_1$ has a positive gap. Consequently, to get to an SP matching without using $R$ or $\overline{R}$, we must reduce this gap in row $j_1$ by using another rotation $R'$ or its dual $\overline{R'}$. However, since $R$ and $R'$ contains the same man $m_{j_1}$, $R'$ is not exposed before $R$ is eliminated. Similarly, $\overline{R}$ and $\overline{R'}$ contain the same woman $w_{j_1}$, so $\overline{R'}$ is not exposed before $\overline{R}$ is eliminated. Hence it is impossible to make the gap of row $j_1$ zero. □

**Algorithm** *SRM*

Input: Roommate instance $I$

1. Construct $f_m(I)$ and $f_w(I)$.

2. Apply the GS algorithm to $\widehat{I} = (f_m(I), f_w(I))$ and let the resulting matching be $M$. If $M$ is not perfect or includes a negative gap row, then exit with NO.

3. While $M$ includes a positive gap row, do:

    (a) Select an arbitrary row with a positive gap, eliminate the rotation including that row (that must exist), and let $M$ be the resulting matching.

    (b) If $M$ is negative-gap, exit with NO.

4. Exit with $M$.

**Theorem 4.** *Algorithm SRM is correct and its running time is $O(n^2)$.*

*Proof.* It is easy to see that constructing $f_m(I)$ and $f_w(I)$ in Step 1 and applying the GS algorithm to $\widehat{I}$ in Step 2 can be done in $O(n^2)$ time each. The while-loop in Step 3 is the process of successively eliminating of rotations, which can also be done in $O(n^2)$ time (see e.g. [3]).

For the correctness, if *SRM* reaches a matching $M$ that is same-position, then it is translated to a stable roommate matching by Lemmas 1 and 2. There are several cases that *SRM* fails.

(1) The GS algorithm provides $M$ that is not perfect. Since all stable matchings have the same set of matched men (e.g., [7]), there is no perfect stable matching for $\widehat{I}$ and there is no stable roommate matching either by Lemmas 1 and 2.

(2) The GS algorithm provides $M$ that includes a negative gap row $l$. Since the GS algorithm provides the man-optimal stable matching, all other stable matchings (if any) must have the same or smaller gap in row $l$. Thus there is no SP stable matching for $\widehat{I}$ and no stable roommate matching either by Lemmas 1 and 2.

(3) The last case is that the current $M$ is negative-gap in Step 3(b). We assume that both men's preference table and women's preference table do not include entries for *unstable pairs*, pairs that can never be a part of any stable matching. This deletion can be done in $O(n^2)$ time. So, elimination of a single rotation moves each partner of the men's side to the right by one position and each partner of the women's side to the left by one position. Note that the previous matching $M'$ is nonnegative-gap and $SRM$ eliminated some rotation $R$ to obtain $M$ from $M'$. There are two cases. One is that $R$ and $\overline{R}$ (that has not yet been eliminated) have a common row. In this case, $R$ precedes $\overline{R}$ in the rotation POSET, meaning that in order to eliminate $\overline{R}$, $R$ must be eliminated before that. Thus the elimination of $R$ is mandatory to reach an SP matching by Lemma 3 and its failing means there is no SP stable matching in this $\widehat{I}$. The other is that $R$ and $\overline{R}$ have no common row. If all the rows of $R$ have positive gaps, the matching after $R$ is eliminated is not negative for the above reason (each gap deceases by one). So, the rows of $R$ have both positive and zero gaps. Then $\overline{R}$ must also have both positive and zero gap rows, namely elimination of either one results in a negative-gap row. Thus by Lemma 3 there is no SP stable matching. □

See Figure 3. In this example, the GS algorithm provides the matching, $M_0$, consisting of the whole first column of the men's table and the whole fourth column of the women's table. Here, entries corresponding to unstable pairs (e.g., the last 3 or $c$ who are originally at the end of the first row) are already deleted. One can see that rotation $(1, f), (6, b), (3, e)$ is exposed in $M_0$ and its dual $(2, a), (5, f), (6, c)$ has an overlapping row 6. Note that $(2, d), (4, c), (5, a)$ is also exposed in $M_0$ and it is again overlapped with its dual $(1, d), (4, e), (3, b)$ in row 4. After eliminating these two rotations, $SRM$ eliminates rotation $(2, c), (4, a), (6, e)$ (which is not overlapped with its dual) and gets to an SP stable matching.

| 1: | 6 | 2 | 4 | 5 | | 1: | $f$ | $b$ | $d$ | $e$ | | $a$: | 6 | 2 | 4 | 5 |
|---|---|---|---|---|---|---|---|---|---|---|---|---|---|---|---|---|
| 2: | 4 | 3 | 1 | 6 | | 2: | $d$ | $c$ | $a$ | $f$ | | $b$: | 4 | 3 | 1 | 6 |
| 3: | 5 | 6 | 2 | 4 | | 3: | $e$ | $f$ | $b$ | $d$ | | $c$: | 5 | 6 | 2 | 4 |
| 4: | 3 | 1 | 5 | 2 | | 4: | $c$ | $a$ | $e$ | $b$ | | $d$: | 3 | 1 | 5 | 2 |
| 5: | 1 | 4 | 6 | 3 | | 5: | $a$ | $d$ | $f$ | $c$ | | $e$: | 1 | 4 | 6 | 3 |
| 6: | 2 | 5 | 3 | 1 | | 6: | $b$ | $e$ | $c$ | $a$ | | $f$: | 2 | 5 | 3 | 1 |

Figure 3: $I$, $f_m(I)$ and $f_w(I)$ of Example 3

The next Example 4 in Figure 4 is an example that $SRM$ fails. Here, again, the result of the GS algorithm corresponds to the first column of the men's table and the fourth column of the women's table. There is only one rotation $(1, b), (4, e), (5, f), (2, c), (3, d), (6, a)$ exposed in $M_0$ and eliminating this yields a stable matching corresponding to the second columns of men's table and the third columns of the women's table. Now, a rotation $(1, e), (5, c), (3, a)$ (whose dual is itself) arises, and eliminating it makes the gaps of rows 1, 3, and 5 negative. Therefore, $SRM$ fails at Step 3(a). Note that there is another rotation $(2, d), (4, f), (6, b)$ but it has the same effect on rows 2, 4, and 6.

| 1: | 2 | 5 | 3 | 6 | 1: | b | e | c | f | a: | 2 | 5 | 3 | 6 |
|---|---|---|---|---|---|---|---|---|---|---|---|---|---|---|
| 2: | 3 | 4 | 6 | 1 | 2: | c | d | f | a | b: | 3 | 4 | 6 | 1 |
| 3: | 4 | 1 | 5 | 2 | 3: | d | a | e | b | c: | 4 | 1 | 5 | 2 |
| 4: | 5 | 6 | 2 | 3 | 4: | e | f | b | c | d: | 5 | 6 | 2 | 3 |
| 5: | 6 | 3 | 1 | 4 | 5: | f | c | a | d | e: | 6 | 3 | 1 | 4 |
| 6: | 1 | 2 | 4 | 5 | 6: | a | b | d | e | f: | 1 | 2 | 4 | 5 |

Figure 4: $I$, $f_m(I)$ and $f_w(I)$ of Example 4

## 4 Simulating Roommate by Marriage

Recall that our second goal is, for a given roommate instance $I$, to obtain a marriage instance $I'$, in polynomial time, such that we can obtain any stable matching of $I$ by solving $I'$. We first give an easy example to show our basic idea, but before that, we give the following corollary of Theorem 4. Let the condition of Lemma 3 be denoted by the *XOR-condition*.

**Corollary 5.** *Let a marriage instance $\widehat{I} = (f_m(I), f_w(I))$ have the rotation POSET $P$. Then there is a one-to-one correspondence between the set of all the SP stable matchings of $\widehat{I}$ and the set of all the closed subsets of $P$ satisfying the XOR-condition.*

*Proof.* By Lemma 3, any SP stable matching corresponds to a closed subset of $P$ satisfying the XOR-condition.

Conversely, let $S$ be a closed subset of $P$ satisfying the XOR-condition, and let $M_S$ be the stable matching corresponding to $S$. We show that $M_S$ is SP. Let $M_0$ and $M_z$ be the man-optimal and woman-optimal stable matchings of $\widehat{I}$, respectively, and $\overline{S}$ be the set of rotations of $P$ that are not in $S$. Note that $M_S$ is obtained from $M_0$ by eliminating all the rotations in $S$. By exchanging the roles of men and women, it is easy to see that $M_S$ can be obtained from $M_z$ by eliminating all the rotations in $\overline{S}$. (Here, rotations are used in the opposite manner, i.e., the rank of matched partner is improved in women's lists and deteriorated in men's lists.) Since our instance $\widehat{I}$ is symmetric, man $m_j$ is matched with woman $w_i$ in $M_0$ iff woman $w_j$ is matched with man $m_i$ in $M_z$, for any $j$. In the process of eliminating rotations from $M_0$, suppose that we eliminate a rotation $R \in S$ and by which, man $m_j$'s partner changes from $w_{i_1}$ to $w_{i_2}$. Then, in the process starting from $M_z$, the rotation $\overline{R} \in \overline{S}$ is eliminated and it changes $w_j$'s partner from $m_{i_1}$ to $m_{i_2}$. This holds for any index $j$ and for any rotation $R$ in $S$. Hence, in $M_S$, the position of $m_j$'s partner and that of $w_j$'s partner is the same, i.e., $M_S$ is SP. □

### 4.1 Easy Example

See Figure 5 for Example 5. This is a roommate instance $I_5$, where and in what follows, we omit $f_m(I_5)$ and $f_w(I_5)$; they are very similar and one can imagine them from $I_5$ easily. However, it is important to note that our all discussion from now on is still done over $f_m(I_5)$ and $f_w(I_5)$. Note that the GS algorithm gives us the whole first column in

```
1:  4  5  6  7  8  3
2:  3  6  8  5  7  4
3:  1  7  5  8  6  2
4:  2  8  7  6  5  1
5:  6  4  2  3  1  7
6:  8  3  4  1  2  5
7:  5  2  1  4  3  8
8:  7  1  3  2  4  6
```

Figure 5: $I_5$ of Example 5

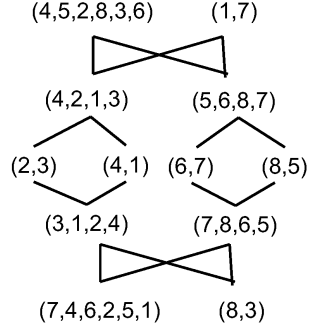

Figure 6: POSET $P_5$

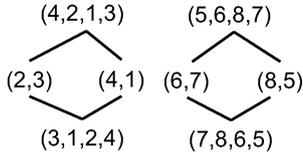

Figure 7: POSET $P'_5$

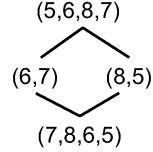

Figure 8: POSET $P''_5$

the men's table and the whole last column for the women's table, as the man-optimal stable matching. Figure 6 shows its rotation POSET $P_5$. Here we also use a simplified description for rotations. For instance, rotation $(1,4),(7,5),(4,2),(6,8),(2,3),(5,6)$ is denoted by just $(4,5,2,8,3,6)$ by enumerating only the second entries. Since each rotation is applicable to some fixed column of the table, there seems to be no confusion. The 12 rotations are decomposed into 6 dual pairs,

$$(1,7) \text{ and } (8,3),$$
$$(4,5,2,8,3,6) \text{ and } (7,4,6,2,5,1),$$
$$(4,2,1,3) \text{ and } (7,8,6,5),$$
$$(5,6,8,7) \text{ and } (3,1,2,4),$$
$$(2,3) \text{ and } (8,5),$$
$$(4,1) \text{ and } (6,7).$$

Observe that $(1,7)$ and $\overline{(1,7)} = (8,3)$ have precedence relation in the POSET $P_5$, and so are $(4,5,2,8,3,6)$ and $\overline{(4,5,2,8,3,6)}$, too. Therefore both $(1,7)$ (that precedes $\overline{(1,7)}$) and $(4,5,2,8,3,6)$ (that precedes $\overline{(4,5,2,8,3,6)}$) must be eliminated by Lemma 3. Now the POSET $P_5$ becomes simpler $P'_5$ in Figure 7 by changing the starting and the ending stable matchings.

The next observation is important: In $P'_5$, no pair of dual rotations has precedence relation in $P'_5$. In other words, we can consider only the right half of $P'_5$, POSET $P''_5$ in Figure 8, which includes only pair-wise independent rotations. The key point is that we can obtain SP matchings by just working over this $P''_5$ without considering the SP condition. To see this, we define function $\sigma_X$ called a *rotation augmenting function with respect to a set of paired rotations* $X$ as follows. Let $S \subseteq X$ be a (possibly empty) set of rotations. Then $\sigma_X(S)$ includes rotation $R$ if $R \in S$ and $\overline{R}$ if $R \notin S$. Note that our current $X$ includes the eight rotations appearing in $P'_5$.

Observe that POSET $P''_5$ has six closed subsets (including the empty set). As one of them, let $S_1 = \{(5,6,8,7),(6,7),(8,5)\}$. Then $\sigma_X(S_1)$ includes those three rotations plus $(4,2,1,3)$. Now one can easily see that this $\sigma_X(S_1)$ is a closed subset of $P'_5$ and obviously satisfies the XOR-condition by the definition of $\sigma_X$. Therefore the matching $\{(1,7),(2,5),(3,8),(4,6),(5,2),(6,4),(7,1),(8,3)\}$, obtained by eliminating all the rotations in $\sigma_X(S_1)$, is SP by Corollary 5. To simplify the notation, we use $[7,5,8,6,2,4,1,3]$ to denote this matching from men's viewpoint, where only the indices of women are described in an increasing order of the matched men's indices. We may use another simplified notation from women's viewpoint; the pairs of the matching are reordered according to women's indices as $\{(7,1),(5,2),(8,3),(6,4),(2,5),(4,6),(1,7),(3,8)\}$, and only the men's indices are given as $[[7,5,8,6,2,4,1,3]]$. (Since this is an SP matching, these [...] and [[...]] must be the same, but not in general.) It then provides the roommate matching $\{(1,7),(2,5),(3,8),(4,6)\}$ by Lemmas 1 and 2. We can similarly verify this property for the remaining five closed subsets of $P''_5$ (e.g., $\sigma_X(\emptyset)$ includes the four left-side rotations of $P'_5$ and provides SP matching $[5,6,7,8,1,2,3,4]$). In this particular example, it is not hard to confirm that each closed subset of $P''_5$ defines a closed subset of $P'_5$ with XOR-condition and so each such subset gives us an SP matching of $(f_m(I_5), f_w(I_5))$ and in turn a stable roommate matching of $I_5$.

We can claim that the converse is also true (again in this particular example). Namely each closed subset of $P'_5$ satisfying the XOR-condition can be mapped to a closed subset of $P''_5$ by reversing $\sigma_X$. It is easy to construct a marriage instance, $I'_5$, of four men and four women, whose rotation POSET is isomorphic to $P''_5$. Thus we were able to reduce the roommate instance $I_5$ to the marriage instance $I'_5$ such that the stable marriage matchings of the latter correspond to the stable roommate matchings of the former in a one-to-one fashion. Our goal is achieved in this example.

## 4.2 Harder Example

The next example, Figure 9, is a popular one included in [3]. The GS algorithm provides the man-optimal stable matching, $[8,4,5,9,7,2,1,10,6,3]$ (from men's viewpoint) or equivalently $[[7,6,10,2,3,9,5,1,4,8]]$ (from women's viewpoint). Figure 10 provides the preference table after the GS algorithm is applied. Note that in our simplified notations of a matching, the former and the latter correspond to the first (leftmost) and the last (rightmost) entries of Figure 10, respectively. As illustrated in Figure 11, its rotation POSET, $P_6$, has two rotations $(8,2)$ and $(2,3,6)$ at the top and their duals at the bottom. Clearly $(8,2)$ precedes $(1,6)$ and $(2,3,6)$ precedes $(9,1,10)$ in $P_6$, so both $(8,2)$ and $(2,3,6)$ must be eliminated to reach an SP matching. After that, we obtain the simplified table, $I'_6$ of Figure 12, corresponding to the matching $[3,4,5,9,7,8,1,10,2,6]$, or equivalently $[[7, 9, 1, 2, 3, 10, 5, 6, 4, 8]]$.

```
 1:   8   2   9   3   6   4   5   7  10
 2:   4   3   8   9   5   1  10   6   7
 3:   5   6   8   2   1   7  10   4   9
 4:  10   7   9   3   1   6   2   5   8
 5:   7   4  10   8   2   6   3   1   9
 6:   2   8   7   3   4  10   1   5   9
 7:   2   1   8   3   5  10   4   6   9
 8:  10   4   2   5   6   7   1   3   9
 9:   6   7   2   5  10   3   4   8   1
10:   3   1   6   5   2   9   8   4   7
```

Figure 9: $I_6$ of Example 6

```
 1:   8   2   3   6   4   7
 2:   4   3   8   9   5   1  10   6
 3:   5   6   2   1   7  10
 4:   9   1   6   2
 5:   7  10   8   2   6   3
 6:   2   8   3   4  10   1   5   9
 7:   1   8   3   5
 8:  10   2   5   6   7   1
 9:   6   2  10   4
10:   3   6   5   2   9   8
```

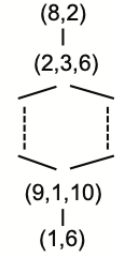

Figure 10: $I_6$ after the GS algorithm applied   Figure 11: POSET $P_6$

Now $P'_6$ in Figure 13 is the rotation POSET for $I'_6$. $P'_6$ includes 10 rotations, which are decomposed into the following five dual pairs:

$(5, 6)$ and $(3, 10)$,
$(3, 4)$ and $(2, 1)$,
$(10, 2)$ and $(8, 9)$,
$(6, 2)$ and $(3, 8)$,
$(9, 1, 5)$ and $(4, 7, 10)$.

It now turns out that none of these dual pairs have precedence relation in $P'_6$. Hence if we choose any one of the top three rotations, its successor rotations are pair-wise independent. Suppose, for example, that we choose $(3, 4)$. Then the POSET consisting of $(3, 4)$ and its successors look like Figure 14, denoted as POSET $P''_6$. Notice that $P''_6$ includes five rotations, i.e., satisfies the XOR condition. In general, such a POSET starting from a single maximal rotation (a rotation that has no preceding ones) is smaller.

One can see there are nine closed subsets of $P''_6$. If each of them corresponds to each closed subset of $P'_6$ in the same way as the previous section, we can use $P''_6$ as the marriage instance corresponding to the roommate instance. Consider for instance a closed subset $S_2 = \{(3,4),(3,8)\}$ of $P''_6$ and let now $X$ be the set of rotations in $P'_6$. We have $\sigma_X(S_2)$ that includes $S_2$ and additional $(5,6)$, $(10,2)$ and $(9,1,5)$ and it turns out that $\sigma_X(S_2)$ is in fact a closed subset of $P'_6$. Unfortunately there is a bad case: Consider another closed subset of $P''_6$, $S_3 = \{(3,4),(3,8),(4,7,10),(8,9)\}$. $\sigma_X(S_3)$ has an additional $(5,6)$, but $S_3 \cup \{(5,6)\}$ is not a closed subset of $P'_6$ because $(8,9)$ is in but $(9,1,5)$ is out.

| | | | | |
|---|---|---|---|---|
| 1: | 3 | 4 | 7 | |
| 2: | 4 | 3 | 8 | 9 |
| 3: | 5 | 6 | 2 | 1 |
| 4: | 9 | 1 | 6 | 2 |
| 5: | 7 | 10 | 8 | 3 |
| 6: | 8 | 3 | 4 | 10 |
| 7: | 1 | 5 | | |
| 8: | 10 | 2 | 5 | 6 |
| 9: | 2 | 10 | 4 | |
| 10: | 6 | 5 | 9 | 8 |

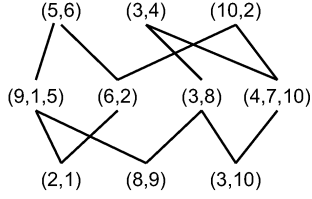

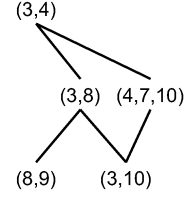

Figure 12: $I'_6$ of Example 6     Figure 13: POSET $P'_6$     Figure 14: POSET $P''_6$

A little fortunately, there are only two such bad cases; the other bad case is that $S_4 = \{(3,4),(3,8),(4,7,10),(8,9),(3,10)\}$. All the other seven cases are ok, i.e., if $S$ is a closed subset of $P''_6$, then $\sigma_X(S)$ is also a closed subset of $P'_6$. The reason why these two closed subsets of $P''_6$ are not closed subsets of $P'_6$ is that both sets include $(8,9)$ and $(4,7,10)$; if $(8,9)$ is in, $(9,1,5)$ must also be in for closedness, but this does not hold since $(9,1,5)$'s dual $(4,7,10)$ is in. We call such a pair as $(8,9)$ and $(4,7,10)$ a *prohibited pair*. (A formal definition of the prohibited pair will be given later in Theorem 9, but to give an idea using the current example, they form a prohibited pair because $(4,7,10)$'s dual $(9,1,5)$ precedes $(8,9)$ in $P'_6$.) One can see that there are seven closed subsets $S$ of $P''_6$, each of which contain at most one of the prohibited pair and $\sigma_X(S)$ is successfully a closed subset of $P'_6$.

To cope with this problem, we use what we call *serialization*. See Figure 15. The idea is to derive two POSETs from $P''_6$, one by deleting $(8,9)$ (and related edges) and the other by deleting $(4,7,10)$ (together with its successor $(3,10)$), as given in Figure 15(a) and (b), respectively. Then we connect them sequentially as in Figure 15(c). Note that its lower part consists of three rotations but their names are not important. What is important is its shape as a graph, i.e., being isomorphic to POSET (b). The new POSET (c), denoted by $P'''_6$, has six closed subsets of POSET (a) and three additional closed subsets corresponding to the three non-empty closed subsets of POSET (b), i.e., $S_0 \cup \{(3',4')\}$, $S_0 \cup \{(3',4'),(3',8')\}$ and $S_0 \cup \{(3',4'),(3',8'),(8',9')\}$, where $S_0$ is the set of all the rotations in the upper part. Notice that the first two of them are basically the same as the two closed subsets $\{(3,4)\}$ and $\{(3,4),(3,8)\}$ already appeared in the upper part; their duplications should be removed. This can be done, for instance, by defining the following function, $\delta$, transferring a set of rotations to another set of rotations: Let $h$ be a function that changes the names of rotations in the lower part to the original names (e.g., $h((3',4')) = (3,4)$). Then $\delta(S) = S$ if $S \subseteq S_0$ and $\delta(S_0 \cup \{r_1, r_2, \ldots, r_k\}) = \{h(r_1), h(r_2), \ldots, h(r_k)\}$ otherwise. For instance $\delta(\{(3,4), (3,8), (4,7,10)\}) = \{(3,4), (3,8), (4,7,10)\}$ and $\delta(\{(3,4), (3,8), (4,7,10), (3,10), (3',4'), (3',8')\}) = \{(3,4), (3,8)\}$. The function $\delta$ is not one-to-one, but it successfully changes the nine closed subsets of $P_6'''$ to the seven closed subsets of $P_6'$ each of which corresponds to each SP matching of $I_6'$.

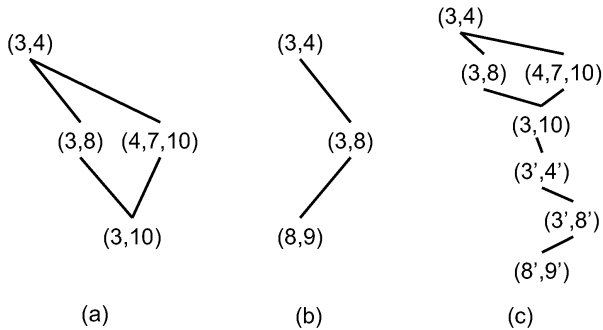

Figure 15: Serialization

|    |    |    |    |    |     |    |    |    |    |     |    |    |    |    |
|----|----|----|----|----|-----|----|----|----|----|-----|----|----|----|----|
|    |    |    |    |    |     |    |    |    |    | 1:  | 3  | 4  | 7  |    |
|    |    |    |    |    |     |    |    |    |    | 2:  | 4  | 3  | 8  |    |
|    |    |    |    |    |     |    |    |    |    | 5:  | 7  | 10 | 3  |    |
|    |    |    |    |    |     |    |    |    |    | 6:  | 8  | 3  | b  |    |
|    |    |    |    |    |     |    |    |    |    | 9:  | 10 | 4  |    |    |
| 1: | 3  | 4  | 7  |    |     |    |    |    |    | a:  | 3' | 10 |    |    |
| 2: | 4  | 3  | 8  |    | 1': | 3' | 4' |    |    | 1': | b  | 3' | 4' |    |
| 5: | 7  | 10 | 3  |    | 2': | 4' | 3' | 8' | 9' | 2': | 4' | 3' | 8' | 9' |
| 6: | 8  | 3  | 10 |    | 6': | 8' | 3' |    |    | 6': | 8' | 3' |    |    |
| 9: | 10 | 4  |    |    | 10':| 9' | 8' |    |    | 10':| 9' | 8' |    |    |

Figure 16: Table $I_6''$     Figure 17: Table $I_6'''$     Figure 18: Table $I_6''''$

Note that there is a general algorithm that constructs a marriage instance $I$ which realizes the rotation POSET $P_6'''$ [4]. However, it is desirable that $I$ naturally reflects the name changing function $h$. In this sense, the following discussion might be a reasonable solution (it is enough to give only the men's table since it is almost automatic to provide an appropriate women's table). We first delete from $I_6'$ all the elements that do not have to do with the rotations in POSET (a). Then we have the table $I_6''$ in Figure 16. Note that the rotation POSET corresponding to $I_6''$ is isomorphic to POSET (a). Similarly, we delete from $I_6'$ all the elements that do not have to do with the rotations in POSET

(b) and obtain $I_6'''$ (Figure 17), whose POSET is isomorphic to POSET (b). Here, the persons are renamed to distinguish from $I_6''$. Finally, we merge $I_6''$ and $I_6'''$ to obtain the final table. To do so, we have to make sure that the rotation $(3, 10)$ must be eliminated before $(3', 4')$. We first simply take the union of two tables. Then, we introduce a man $a$ and a woman $b$, and modify the preference lists of 6 (corresponding to the rotation $(3, 10)$) and $1'$ (corresponding to the rotation $(3', 4')$) as in Table $I_6''''$ (Figure 18). Now the rotation $(3, 10)$ in POSET (a) is extended to $(3, b, 3', 10)$ which inhibits exposure of $(3', 4')$ until it is eliminated. This marriage instance $I_6''''$ has nine stable matchings, but they are converted to seven SP matchings of $I_6'$ by $\delta$, which are in turn converted to the seven roommate solutions of the original $I$ in a obvious way.

The following may be worth mentioning: Recall that POSET (b) is just a simple path. So, if we take only the rotation in the prohibited pair, $(8, 9)$, and connect only this one after POSET (a), then we do not have to worry about the above duplication of closed subsets and the combined POSET ($(3', 4')$ and $(3', 8')$ removed from (c)) has seven stable matchings. Thus we could achieve the one-to-one correspondence as the previous easy example. Unfortunately, however, this approach is unlikely to work if the POSET (b) is complex as a graph, i.e., to construct a POSET that exactly realizes only the closed subsets involving the rotation in the prohibited pair seems hard (see the next subsection).

## 4.3 Reduction Algorithm

In the previous subsection, there are two key ingredients in the procedure of converting a roommate instance ($I_6$ in the example) to the final marriage instance ($I_6''$). (Recall that the latter is a "pure" marriage instance for which we do not have to consider any restrictions such as prohibited pairs or SP, although its stable matching should be mapped by functions $\delta$ and $\sigma_X$ to obtain roommate matchings.) One is the conversion of POSET from $P_6$ to $P_6''$ and the other from $P_6''$ to POSET (c) of Figure 15. We first consider the former for which we need the following preparation. A rotation POSET is said to be *irreducible* if there is no rotation $R$ such that (i) $R$ precedes $\overline{R}$ or (ii) $R$ precedes both $R'$ and $\overline{R'}$ for some $R'$. A rotation POSET can be changed to an irreducible one by removing all rotations $R$ satisfying (i) or (ii) (and their duals $\overline{R}$) and we can easily show that an irreducible POSET preserves all SP matchings of the original POSET by Lemma 3. Fix an arbitrary irreducible POSET. Define a *maximal rotation* as a rotation that has no preceding ones in the POSET. For a maximal rotation $R$, $\pi(R)$ is defined as the rotation POSET consisting of $R$ and its all successors. Because of irreducibility, $\pi(R)$ includes only pair-wise disjoint rotations, denoted by $V(\pi(R))$. It then turns out that all the duals of $V(\pi(R))$ constitute a "dual POSET", $\overline{\pi(R)}$, due to the symmetric structure of $\widehat{I} = (f_m(I), f_w(I))$. For example, $P_6'$ is irreducible and $\pi((3, 4))$ is $P_6''$. $\overline{\pi((3, 4))}$ consists of rotations $(2, 1)$, $(6, 2)$, $(9, 1, 5)$, $(5, 6)$ and $(10, 2)$. We are now ready to give our first conversion algorithm.

**Algorithm** *HalfCut*
Input: Roommate instance $I$

1. Run *SRM* to check if $\widehat{I} = (f_m(I), f_w(I))$ has at least one SP stable matching. If *SRM* fails, exit with NO.

2. Compute the rotation POSET $P$ of $\widehat{I}$ using the algorithm in [3] and its irreducible version $P'$ as described above. Let $Q = P'$ and $A = \emptyset$.

3. While $Q$ is not empty, do

   (a) Select a maximal rotation $R$ in $Q$, let $A = A \cup V(\pi(R))$, and remove $\pi(R)$ and $\overline{\pi(R)}$ from $Q$.

4. Exit with the sub-POSET of $P'$ induced by $A$.

**Lemma 6.** *Let $P$ be the rotation POSET of $\widehat{I} = (f_m(I), f_w(I))$ and $P'$ be its irreducible subPOSET. Then* HalfCut *provides a POSET $P''$ including one half rotations of $P'$ which are pair-wise disjoint.*

*Proof.* In Step 3(a) of *HalfCut*, let $T$ be the set of rotations added to $A$. Then $T$ includes only pair-wise disjoint rotations and all of their duals are discarded from $Q$ as $\overline{\pi(R)}$. So if $R$ is in $A$, $\overline{R}$ is not in $A$ at any moment of the execution. This is enough to claim the lemma. □

Now we have POSET $P''$. Its size is exactly one half $P'$, the irreducible subPOSET of the original $P$, but the next lemma shows that $P''$ preserves enough information about all the closed subsets of $P'$ satisfying the XOR-condition.

**Lemma 7.** *Let $S$ be a closed subset of $P'$ satisfying the XOR-condition. Then $S \cap P''$ is a closed subset of $P''$.*

*Proof.* Suppose not. Then there are rotations $R_1 \in P'' \setminus S$ and $R_2 \in S \cap P''$ such that $R_1$ precedes $R_2$ in $P''$. But since $P''$ is a subPOSET of $P'$ (see *HalfCut*), $R_1$ must be in $P'$. Since $R_1$ is in $P''$, its absence in $S \cap P''$ means that it is not in $S$. Then it contradicts closedness of $S$ in $P'$. □

If the converse of Lemma 7 is also true, namely if all closed subsets of $P''$ are also closed subsets of $P'$ after augmented by $\sigma_X$, we are done. This actually happens in the easy example. However, as shown in the harder example, the converse is not true in general, i.e., some closed subset $S''$ in $P''$ may not be closed in $P'$ after augmented to $\sigma_X(S'')$, where $X$ is the set of rotations in $P'$. The following lemma shows a condition for this to happen.

**Lemma 8.** *Let $S''$ be a closed subset of $P''$ and suppose that there are rotations $R_1$ and $R_2$ such that $R_1$ precedes $R_2$ in $P'$ and $R_1 \notin \sigma_X(S'')$ but $R_2 \in \sigma_X(S'')$. Then $R_1 \notin P''$ and $R_2 \in P''$.*

*Proof.* Let $R_1$ and $R_2$ satisfy the condition of the lemma. Then the predicate of the conditional part of the lemma, $R_1 \notin \sigma_X(S'')$ but $R_2 \in \sigma_X(S'')$, cannot be true if $R_1$ and $R_2$ are both in $P''$ since $S''$ is its closed subset, nor if neither of $R_1$ and $R_2$ is in $P''$ because of the symmetric structure of $P'$ (the order of $R_1$ and $R_2$ is opposite to the order of $\overline{R_1}$ and $\overline{R_2}$). If $R_1$ is in $P''$ then $R_2$ must also be in $P''$ since $R_1$'s all successors are also added in $P''$ by *HalfCut* (if $R_2$ has been deleted, $R_1$ has also been deleted at that moment). Therefore the remaining only one possibility is that $R_1 \notin P''$ and $R_2 \in P''$. □

Recall that if $R_1$ is not in $P''$, $\overline{R_1}$ must be in $P''$. Thus the bad case of Lemma 8 happens only if there are two specific rotations (such as $\overline{R_1}$ and $R_2$ above) are in $S''$. They form exactly a prohibited pair introduced in the previous section. In other words, if a closed subset of $P''$ does not contain a prohibited pair, it is a closed subset of $P'$ after mapped by $\sigma_X$. We thus have the following theorem.

**Theorem 9.** *Let $I$, $P$, $P'$ and $P''$ be the original roommate instance, the rotation POSET of $\widehat{I} = (f_m(I), f_w(I))$, its irreducible one and the output of* HalfCut, *respectively. Also rotations $R_1$ and $R_2$ in $P''$ are called a prohibited pair if $\overline{R_2}$ precedes $R_1$ in $P'$. Then a closed subset $S$ of $P''$ does not include both rotations of any prohibited pair iff $\sigma_X(S)$ is a closed subset of $P'$.*

## 4.4 Only One Prohibited Pair

In this paper, we discuss the case that there is only one prohibited pair (as in the harder example in Sec. 4.2), which may look extremely restricted but we believe is still nontrivial.

Once again recall that our procedure, for a given roommate instance $I$, computes the rotation POSET $P$ of $\widehat{I} = (f_m(I), f_w(I))$, makes it irreducible one $P'$, and use *HalfCut* to obtain the half-size $P''$ comprising pair-wise disjoint rotations. Suppose $P''$ looks like $P_7$ illustrated in Figure 19, which has a single prohibited pair $(j, q)$. We partition $P_7$ into three parts, A to C: A is the part including rotations that are not preceded by any prohibited-pair rotation. Part B is the successors of one of the prohibited-pair rotations, namely $j$, and C for the other, $q$. There can be edges from one part to another like from $c$ to $l$ and from $p$ to $q$, but we can assume that there is no edge from B to C or from C to B. Since it is impossible to eliminate both B and C, those edges and related rotations can be removed. For instance, if there is an edge from $s$ to $m$, we may delete $m$ and its successors.

Figure 20 is the POSET $P_8$ such that rotation $j$ (one of the prohibited-pair rotation) and its successors are deleted, and Figure 21 is the POSET $P_9$ similarly constructed for $q$. Figure 22 is the sequential connection of $P_9$ and $P_8$. Note that the rotation $a'$, a maximal rotation of $P_8$, is connected with each rotation of $P_9$ that has no successor, and hence $a'$ becomes a successors of all the rotations of $P_9$. Let U be the upper part, L be the lower part, and $S_U$ be the set of all the rotations in U. Recall that we have designed a function $\delta$ (see Sec. 4.2) which maps a closed subset of $P_{10}$ to that of $P_7$. If $S$ includes only rotations in U ($S \subseteq S_U$), $\delta(S) = S$. Otherwise, note that a closed subset including rotations in L must include $S_U$ also. In this case, $\delta$ removes all the rotations in $S_U$ and renames the remaining rotations in L to the original names. This gives us a closed subset of $P_7$ and then we can use $\sigma_X$ to compute closed subsets of $P'$ each of which corresponds to each SP matching of $\widehat{I}$.

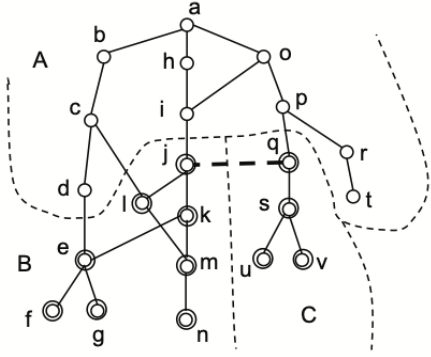 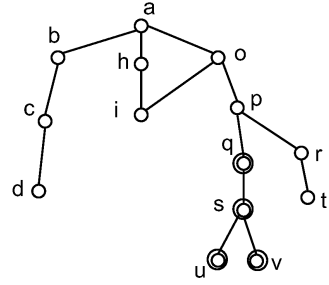

Figure 19: POSET $P_7$        Figure 20: POSET $P_8$

Let's see how this approach works: Suppose that $S$ is a closed subset of $P$ corresponding to an SP matching of $\widehat{I}$. Then $S'$, which is obtained by removing rotations in $P \setminus P'$, is also a closed subset of $P'$ corresponding to the same SP matching. By Lemma 7, $S'$ implies $S''$ that is a closed subset of $P_7$ and this $S''$ does not include both of a prohibited pair by Theorem 9. For a mapping from $S''$ to a closed subset of $P_{10}$, there are a couple of cases: (i) $S''$ consists of only rotations in A. Then it corresponds to two closed subsets of $P_{10}$, one is the copy of $S''$ in U and the other is the whole $S_U$ plus the copy of $S''$ in L. (ii) $S''$ consists of rotations in A and B. Then it corresponds to the copy of $S''$ in U. (iii) $S''$ consists of rotations in A and C. Then it corresponds to the whole $S_U$ plus the copy of $S''$ in L. Thus every SP matching can be mapped to a closed subset of $P_{10}$.

For the converse, note that a series of translations in the above paragraph is almost one-to-one. Only the exception is Case (i) of the mapping from a closed subset $S''$ (containing neither of the prohibited pair) of $P_7$ to a closed subset of $P_{10}$. It is associated with two closed subsets of $P_{10}$ to $S''$, but $\delta$ successfully maps these two to $S''$. (Note that Cases (ii) and (iii) are just the inverse of $\delta$.) Thus every closed subset of $P_{10}$ eventually implies an SP matching of $\widehat{I}$.

It is clear that the whole procedure runs in polynomial time. Formally speaking, we need to change $P_{10}$ to the equivalent preference-list table of the marriage problem. However, it is known to be possible in general [4] and some useful consideration for our specific situation is given in Sec. 4.2. Details may be omitted.

## 5 Final Remarks

An obvious question is what happens if there are two or more prohibited pairs. Our approach, serialization, seems to work for two or more prohibited pairs. For instance, if there are two such pairs, we just have to consider four subPOSETs instead of two and we can connect them sequentially in the same way as before. (If one rotation of one pair precedes another rotation of the other pair, the construction of the subPOSETs is a little more complicated but not much.) However the number of subPOSETs increases exponentially as the number of pairs grows and hence we can no longer claim the efficiency of the approach. In other words, the number of prohibited pairs can be regarded as a kind of measure suggesting how harder the roommates problem is compared to the marriage problem. It should also be noted that the number of prohibited pairs depends on how to select a maximal rotation in Step 3(a) of *HalfCut*. Investigation of these two issues must be an important goal for future research.

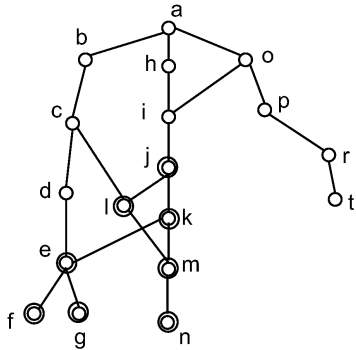 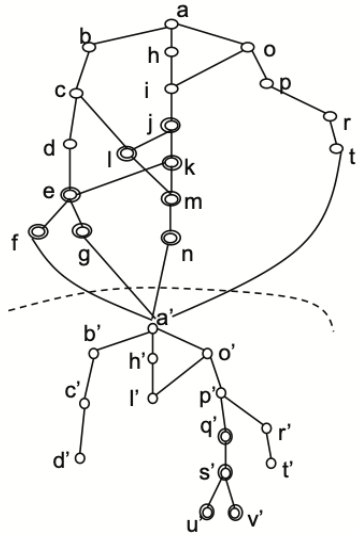

Figure 21: POSET $P_9$      Figure 22: POSET $P_{10}$